\documentclass{aa-package7.0/aa}

\usepackage{natbib}
\bibpunct{(}{)}{;}{a}{}{,}

\usepackage{amsmath}
\usepackage{amsfonts}
\usepackage{amssymb}
\usepackage{wasysym}
\usepackage{textcomp}

\usepackage{booktabs}
\usepackage{graphicx}
\usepackage{color}

\newcommand*{\eqq}[1]{\lq\lq#1\rq\rq}
\newcommand*{\eq}[1]{\lq#1\rq{}}

\newcommand*{\mass}{\mathcal{M}}
\newcommand*{\grad}{\ensuremath{^\circ}}
\newcommand*{\diff}[1]{\text{d}#1}

\newcommand*{\yr}{\,\hbox{yr}}
\newcommand*{\kelvin}{\,\hbox{K}}
\newcommand*{\AU}{\,\hbox{AU}}
\newcommand*{\Jy}{\,\hbox{Jy}}
\newcommand*{\mJy}{\,\hbox{mJy}}
\newcommand*{\mum}{\,\hbox{\textmu m}}
\newcommand*{\mm}{\,\hbox{mm}}

\newcommand*{\m}{\,\hbox{m}}
\newcommand*{\km}{\,\hbox{km}}
\newcommand*{\kmpersec}{\,\hbox{km\,s}\ensuremath{^{-1}}}
\newcommand*{\gramperccm}{\,\hbox{g\,cm}\ensuremath{^{-3}}}
\newcommand*{\ergperg}{\,\hbox{erg\,g}\ensuremath{^{-1}}}

\begin{document}
\newcommand*{\avk}[1]{{\bf [#1 -- AVK.]}}
\newcommand*{\mr} [1]{\color{black}\color{blue}[MR: #1]\color{black}}
\newcommand*{\ccs}[1]{{\bf [#1 -- CCS.]}}
\newcommand*{\rev}[1]{\textbf{#1}}
\newcommand*{\delete} [1]{\color{black}\color{blue}[#1 -- deleted]\color{black}}


\title{The cold origin of the warm dust around $\varepsilon$~Eridani}

\author{M.~Reidemeister\inst{1}
        \and
        A.~V.~Krivov\inst{1}
        \and
        C.~C.~Stark\inst{2}
        \and
        J.-C.~Augereau\inst{3}
        \and
        T.~L\"ohne\inst{1}
        \and
        S.~M\"uller\inst{1}
       }
\offprints{Martin Reidemeister\\ \email{martin.reidemeister@astro.uni-jena.de}}
\institute{Astrophysikalisches Institut, Friedrich-Schiller-Universit\"at Jena,
           Schillerg\"a{\ss}chen~2--3, 07745 Jena, Germany
           \and
           Department of Physics, University of Maryland, Box 197, 082 Regents Drive, College Park, MD 20742-4111, USA
            \and
           Laboratoire d'Astrophysique de Grenoble, CNRS UMR 5571, Universit\'{e} Joseph Fourier, Grenoble, France
          }

\date{Received {\em 2.\,7.\,2010}; accepted {\em 29.\,10.\,2010}}

\abstract
{
  The nearby K2~V star $\varepsilon$~Eridani hosts one known inner planet,
  an outer Kuiper belt analog, and an inner disk of warm dust.
  Spitzer/IRS measurements indicate that the warm dust is present at distances as close as
  a few AU from the star. Its origin is puzzling, since an \eqq{asteroid belt}
  that could produce this dust would be unstable because of the known inner planet.
}
{
  Here we test a hypothesis that the observed warm dust
  is generated by collisions in the outer belt and is transported inward by
  Poynting-Robertson drag and strong stellar winds.
}
{
  We simulated a steady-state distribution of dust particles outside $10\AU$
  with a collisional code and in the inner region ($r < 10\AU$) with single-particle
  numerical integrations.
  By assuming homogeneous spherical dust grains
  composed of ice and astrosilicate, we calculated the thermal emission of
  the dust and compared it with observations. We investigated two different
  orbital configurations for the inner planet inferred from radial velocity
  measurements, one with a highly eccentric orbit of $e = 0.7$ and another one
  with a moderate eccentricity of $e = 0.25$.
  We also produced a simulation without a planet.
}
{
  Our models can reproduce the shape and magnitude of the observed
  spectral energy distribution from mid-infrared to submillimeter wavelengths,
  as well as the Spitzer/MIPS radial brightness profiles.
  The best-fit dust composition includes both water ice and silicates.
  The results are similar for the two possible planetary orbits and without a planet.
}
{
  The observed warm dust in the $\varepsilon$~Eridani system can indeed stem from the
  outer \eqq{Kuiper belt} and be transported inward by Poynting-Robertson and stellar
  wind drag. The inner planet has little effect on the distribution of dust, so that the planetary
  orbit could not be constrained. Reasonable agreement between the model and observations
  can only be achieved by relaxing the assumption of purely silicate dust and assuming a
  mixture of silicate and water ice in comparable amounts.
}

\keywords{planetary systems: formation --
          circumstellar matter --
          stars: individual: $\varepsilon$~Eridani --
          planet-disk interactions --
          zodiacal dust
         }

\authorrunning{Reidemeister~et~al.}
\titlerunning{$\varepsilon$~Eridani}

\maketitle


\section{Introduction} \label{sec: introduction}

\begin{figure}
  \centering
  \includegraphics[width=0.48\textwidth]{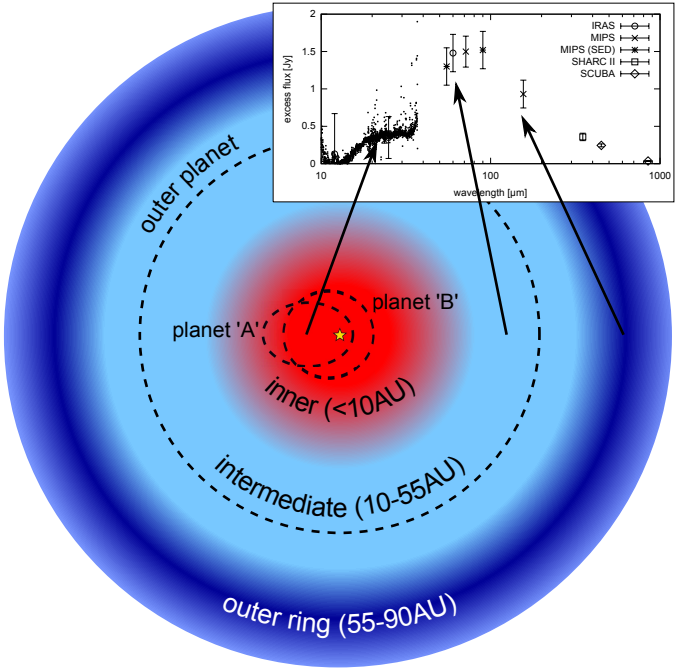}
  \caption{
  A schematic view of the $\varepsilon$~Eridani system's architecture.
  The outer ring is the region where the dust is produced by parent planetesimals;
  the intermediate zone is the one where it is transported inward by drag forces,
  possibly interacting with a presumed outer planet;
  transport continues through the inner region where dust interacts with
  the known inner planet.
  Two possible orbits of the inner planet are shown.
  The outer part of the sketch ($>10\AU$) is not to scale.
  {\em Inset:}
  The observed SED. The IRS spectrum (dots) stems from dust in the inner region
  and exhibits a characteristic \eqq{plateau} (\eqq{shoulder}) at $\lambda \approx 20$--$30\mum$.
  The main part of the SED with a maximum at $\lambda \approx 70$--$80\mum$,
  well probed by several photometry points (symbols with error bars),
  derives from the outer and intermediate regions.
  }
  \label{fig: sketch}
\end{figure}

The nearby ($\sim$ 3.2~pc) K2~V star $\varepsilon$~Eridani
(HD~22049, HIP~16537, HR~1084), with an age of
$\la 1$~Gyr \citep{saffe-et-al-2005, difolco-et-al-2004, song-et-al-2000,
soderblom-dappen-1989}, has a ring of cold dust at $\sim 65\AU$ seen in
resolved sub-millimeter images \citep{greaves-et-al-1998,
greaves-et-al-2005}, which encompasses an inner disk of warm dust revealed
by Spitzer/MIPS \citep{backman-et-al-2009}. The star is orbited by
a radial velocity planet \citep{hatzes-et-al-2000} with a semimajor axis of
3.4\AU. Another outer planet may orbit near $\sim 40\AU$,
producing the inner cavity and clumpy structure in the outer ring
\citep{liou-zook-1999, ozernoy-et-al-2000, quillen-thorndike-2002,
deller-maddison-2005}. The excess emission at $\lambda \gtrsim 15\mum$ in
a Spitzer/IRS-spectrum \citep{backman-et-al-2009} indicates that there is
warm dust close to the star, at a few AU
(inset in Fig.~\ref{fig: sketch}). Its origin is unknown, as an inner
\eqq{asteroid belt} that could produce this dust would be dynamically unstable
because of the known inner planet \citep{brogi-et-al-2009}.

Here, we check the possibility that the source of the warm dust is the outer
ring from which dust grains could be transported inward by
Poynting-Robertson drag and stellar wind.
The importance of the latter for debris disks around late-type, low-mass stars
was first pointed out by \citet{plavchan-et-al-2005}, and it is well known that
$\varepsilon$~Eridani does have strong winds \citep{wood-et-al-2002}.

It is convenient to divide the entire system into three regions:
the outer \eqq{Kuiper belt} (55--90 AU),
the intermediate zone (10--55 AU),
and the inner region (inside 10 AU),
see Fig.~\ref{fig: sketch}.
In Sect.~\ref{sec: setup} we describe the modeling setup.
In Sect.~\ref{sec: outer_interm} we model the dust production in the outer ring
and its transport through the intermediate region.
Section~\ref{sec: inner} describes simulations of the inner system.
Section~\ref{sec: SED} presents the spectral energy distribution (SED) modeling
and provides an additional check for
connection between the inner system and the outer parent ring
(Sect.~\ref{ssec: connection}).
Section~\ref{sec: profiles} focuses on the surface brightness profiles.
Section~\ref{sec: conclusions} contains our conclusions.

\section{Model setup} \label{sec: setup}
\subsection{Method} 

Our model includes gravitational forces from the star and an inner planet, radiation
and stellar wind pressure, drag forces induced by both stellar photons and
stellar wind particles \citep{burns-et-al-1979}, as well as collisions.
Dust production in the outer region and dust transport through the intermediate region
are modeled with a statistical collisional code.
To study the dust in the inner region, we need to handle dust interactions with the inner
planet. These cannot be treated by the collisional code, so we model the
inner region by collisionless numerical integration.

\subsection{Stellar properties} \label{ssec: star}

We assumed a stellar mass of $\mass_\star = 0.83 \mass_{\odot}$ \citep{benedict-et-al-2006}
and a luminosity of $L_\star = 0.32 L_{\odot}$ \citep{difolco-et-al-2007}.
For the stellar spectrum we used a NextGen model
\citep{hauschildt-et-al-1999} with an effective temperature of $5200\kelvin$,
$\log{g} = 4.5$, solar metallicity, and stellar radius $R_\star = 0.735R_\odot$
\citep{difolco-et-al-2007}.

\subsection{Dust grain properties} \label{ssec: grain sizes}

The knee in the IRS spectrum at $\sim 20\mum$
(inset in Fig.~\ref{fig: sketch})
is reminiscent of a classical silicate feature.
Since the exact composition of those silicates is not known,
we have chosen astronomical silicate
\citep{laor-draine-1993} ($\rho_\mathrm{d} = 3.5\gramperccm$).
On the other hand, by analogy with the surface composition of Kuiper belt objects
in the solar system \citep[e.g.,][]{Barucci-et-al-2008},
we may expect many additional species such as ices and organic
solids. In particular, it is natural to expect water ice
to be present, especially given that the source of dust is a "Kuiper belt''
located very far from the star ($\sim 55$--$90\AU$),
and the star itself has a late spectral class.
Accordingly, we also
tried homogeneous mixtures of astrosilicate with
50\% and 70\% volume fraction of water ice
\citep{li-greenberg-1998-03} ($\rho_\mathrm{d} = 1.2\gramperccm$).
The bulk density of these ice-silicate mixtures is
$\rho = 2.35\gramperccm$
and $\rho = 1.89\gramperccm$, respectively.
The optical constants of the mixtures were calculated by effective medium theory
with the Bruggeman mixing rule.
In all three cases (pure astrosilicate, two ice-silicate mixtures)
the dust grains were assumed to be compact spheres.

\subsection{Radiation pressure}

\begin{figure}[t!]
  \centering
  \includegraphics[width=0.48\textwidth]{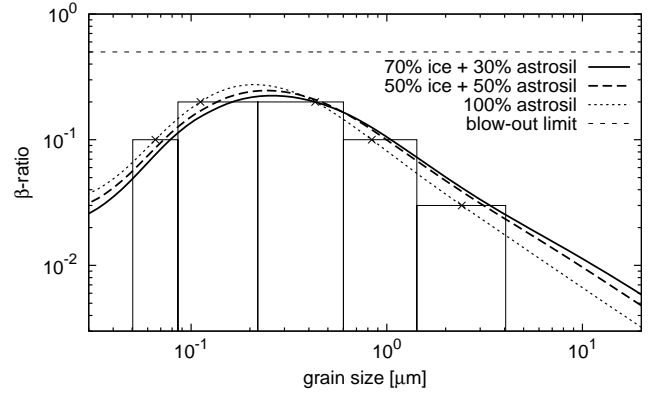}
  \caption{
  The $\beta$ ratio of two silicate -- water ice mixtures,
  compared to a pure silicate composition,
  as a function of size for $\varepsilon$~Eri.
  The bars show the size bins used in the inner disk simulations
  (Sect.~\ref{sec: inner}).
  The dashed horizontal line shows the dynamical blowout limit of $\beta = 0.5$.}
  \label{fig: beta-size}
\end{figure}

Using the optical constants and adopting Mie theory,
the radiation pressure efficiency $Q_\mathrm{rp}$,
averaged over the stellar spectrum, was calculated as a function of size~$s$
\citep{burns-et-al-1979, gustafson-1994}.
We then computed the radiation pressure to gravity ratio, $\beta$
(Fig.~\ref{fig: beta-size}).
The resulting $\beta(s)$ was utilized to compute the direct radiation pressure
and Poynting-Robertson forces.

\subsection{Stellar wind}

The stellar wind was included by a factor $\beta_\mathrm{sw}/\beta$,
which is the ratio of stellar wind pressure to radiation pressure:
\begin{align}
  \frac{\beta_\mathrm{sw}}{\beta} =
  \frac{F_\mathrm{sw}}{F_\mathrm{rp}} =
  \frac{\dot{\mass}_{\star} v_\mathrm{sw} c}{L_{\star}}\frac{Q_\mathrm{sw}}{Q_\mathrm{rp}} \;,
\end{align}
where $Q_\mathrm{sw}$ is the efficiency factor for stellar wind pressure
\citep{burns-et-al-1979, gustafson-1994}. We adopted $Q_\mathrm{sw} = 1$.
Assuming a stellar wind velocity equal to the average solar wind velocity,
$v_\mathrm{sw} \approx 400\kmpersec$, and using a mass-loss rate of
$\dot{\mass}_{\star} \approx 30\dot{\mass}_{\odot}$ \citep{wood-et-al-2002},
we get $\beta_\mathrm{sw} = 3.69 \times 10^{-2}\beta$.
In this estimate (but not in the simulations), we set
$Q_\mathrm{rp}$ to unity.
Thus direct stellar wind pressure is approximately 27 times weaker than
radiation pressure. However, the stellar wind drag is
$(\beta_\mathrm{sw} / \beta) (c / v_\mathrm{sw}) \approx 28$ times
stronger than radiation (Poynting-Roberston) drag.

\subsection{Sublimation}
\label{ssec: sublimation}

\begin{figure}[th!]
  \centering
  \includegraphics[width=0.48\textwidth]{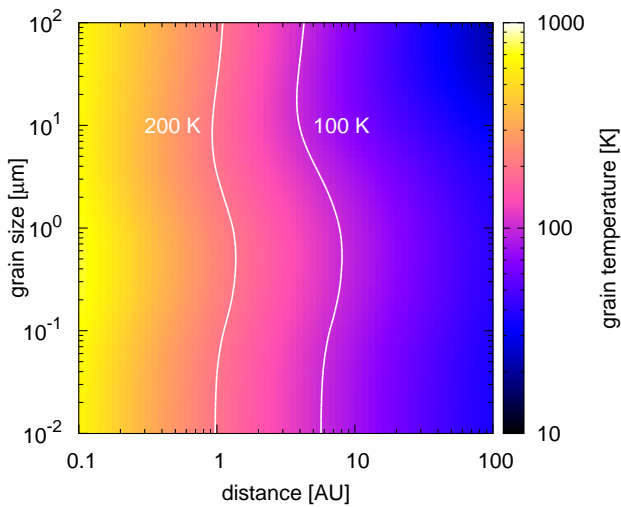}
  \caption{Equilibrium grain temperature of an 70\% ice and 30\%
  astronomical silicate composition depending on grain size $s$ and distance
  to the star $r$. The results for the 50\% ice -- 50\% silicate mixture are similar.
  The wavy shape of the isotherms is caused by the dependence of the absorption
  efficiency on the grain size \citep[as explained in][see their Figs.~2 and 6]{krivov-et-al-2008}.
  }
    \label{fig: temperature}
\end{figure}

Using an ice-silicate mixture for dust raises a question
whether and where the icy portion of the dust grains
will suffer sublimation.
The ice sublimation temperature of $\sim 100\kelvin$
\citep[e.g.][]{moro-martin-malhotra-2002,kobayashi-et-al-2008} is reached at
$\approx 10\AU$ (Fig.~\ref{fig: temperature}).
\citet{mukai-fechtig-1983} proposed that fluffy and nearly
homogeneous grains of ice and silicates can produce a core of silicate after
sublimation of ice.
Assuming a grain with a volatile icy mantle , instead of a homogeneous sphere of 
ice and silicates, would lead to the same result.
In both cases, the silicate cores left after ice sublimation
continue to drift further inward.
Within the ice sublimation distance, the optical depth would decrease by the volume
fraction of the refractories \citep[see][their Eq.~(35)]{kobayashi-et-al-2008}.
The shape of the size distribution is not affected by sublimation,
as long as the volume fraction of ice is independent of the grain size.

Thus we expect the entire dust disk to consist of an inner silicate disk
($r \la 10\AU$) and an outer disk of an ice-silicate composition.
Accordingly, in integrations described in Sect.~\ref{sec: inner}
we assumed dust inside $10\AU$ to consist of pure astrosilicate.
Outside $10\AU$, we tried all three dust compositions described above.

We  finally note that a ring due to sublimation as described in
\citet{kobayashi-et-al-2008,kobayashi-et-al-2009} is not expected,
because such a ring can only be produced
by particles with high $\beta$ ratios ($\beta \ga 0.1$--0.3).
Such $\beta$ ratios are barely reached
in the $\varepsilon$~Eri system even for pure silicate dust.

\begin{figure*}[htb!]
  \centering
  \includegraphics[width=0.48\textwidth]{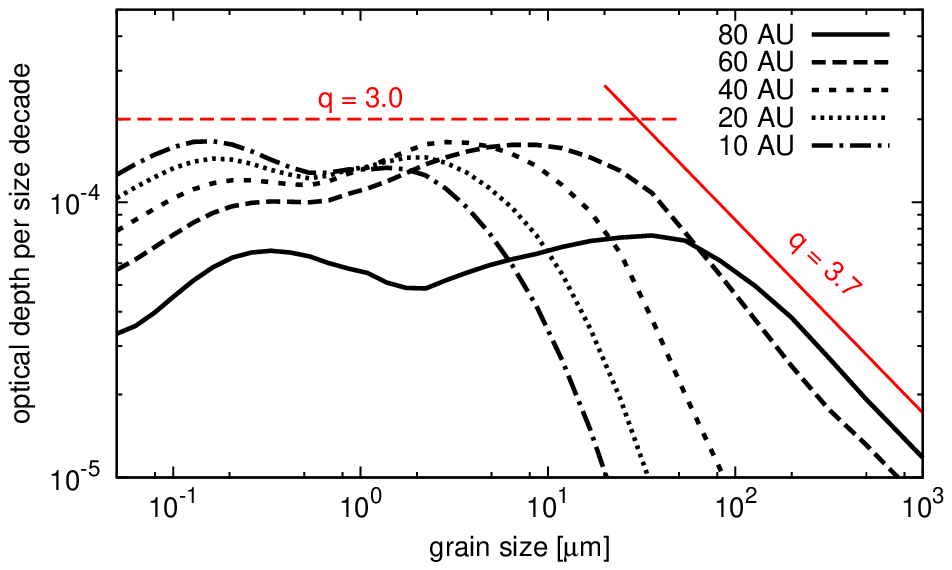}
  \includegraphics[width=0.48\textwidth]{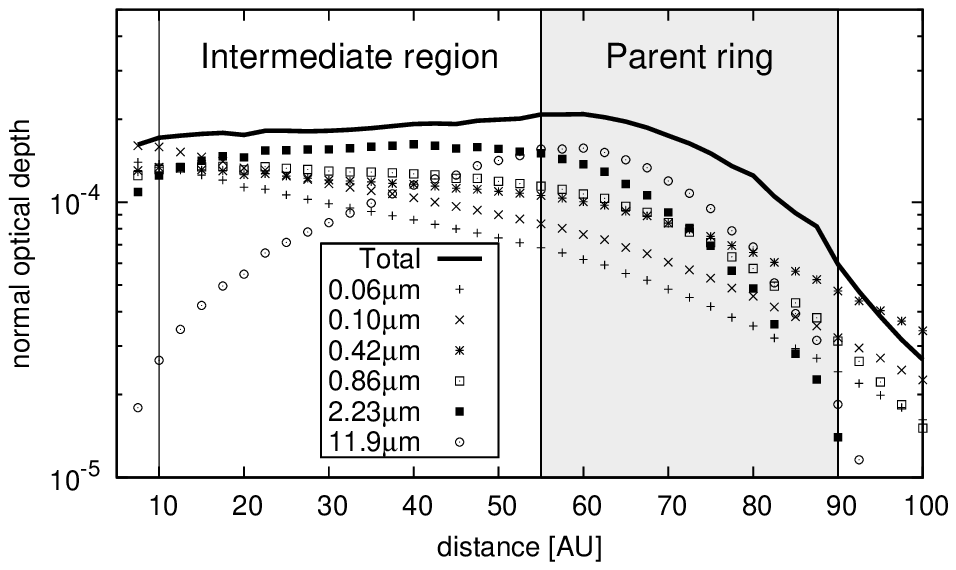}
  \caption{
  {\em Left:} Size distribution at different
  distances in a disk composed of 70\% ice and 30\% silicate, simulated with \emph{ACE}.
  Curves depict the normal geometrical optical depth per size decade, as a function of size.
  The straight lines show power laws with $q=3.0$ and $3.7$ for comparison.
  (Their height is arbitrary.)
  {\em Right:}
  Radial profiles of the optical depth.
  Symbols: partial contributions to the optical depth by several selected sizes.
  Each contribution is normalized to the unit size decade around that size.
  Solid line: the total optical depth of the disk.
  }
  \label{fig: distribution}
\end{figure*}

\section{Dust in the outer and intermediate region} \label{sec: outer_interm}

\subsection{Model}

We used our statistical collisional code \emph{ACE}
(\emph{Analysis of Collisional Evolution})
\citep{krivov-et-al-2005,krivov-et-al-2006, loehne-et-al-2008}
to model the collisional disk beyond 10\AU{}.
This includes the parent ring near 65\AU{} \citep{greaves-et-al-2005}
and the intermediate region inside
it.
The \emph{ACE} simulation provides us with a rotationally-symmetric
steady-state dust size distribution at 10\AU, which we 
use later
as input
for our non-collisional models of the inner disk (see Sect.~\ref{sec: inner}).
The code is not 
able to treat 
planetary perturbations.
Thus we neglect the presumed outer planet, but will qualitatively
discuss its possible influence in Sects.~\ref{ssec: results} and \ref{sec: SED}.

We made three \emph{ACE} runs, assuming the dust composition to be
pure astrosilicate, a mixture of ice and astrosil in equal parts,
and a mixture of 70\% ice and 30\% astrosil, as described above.
In handling the collisions, the critical specific energy
for fragmentation and dispersal $Q_\mathrm{D}^*$
is calculated by the sum of two power laws,
\begin{equation}
  Q_\mathrm{D}^* = A_\mathrm{s} \left(\frac{s}{1\m}\right)^{3b_\mathrm{s}}
                 + A_\mathrm{g} \left(\frac{s}{1\km}\right)^{3b_\mathrm{g}} ,
\end{equation}
where the first and the second terms account for the strength and the gravity regimes,
respectively.
We took the values
$A_\mathrm{s} = 10^7 \ergperg$,
$A_\mathrm{g} = 10^7 \ergperg$,
$3b_\mathrm{s} = -0.37$, and $3b_\mathrm{g} = 1.38$
\citep[cf.][]{benz-asphaug-1999}.
In all \emph{ACE} runs, we assumed that the outer ring has a uniform surface density
between $55$ and $90\AU$ and that the eccentricities of parent planetesimals
range between 0.0 and 0.05, and
arbitrarily set the semi-opening angle of the disk to $3^\circ$.
Next, we assumed that the \emph{initial} size distribution
index of solids from km-sized planetesimals down to dust is $q \sim 3.6$
\citep[see, e.g.][for a justification of this choice]{loehne-et-al-2008}.
The current dust mass was set to $\mass(s<1\mm) \approx 10^{-3}\mass_{\oplus}$,
based on previous estimates from sub-millimeter images
\citep{greaves-et-al-1998,greaves-et-al-2005}.

\subsection{Results} \label{ssec: results}

Figure~\ref{fig: distribution} (left) shows the resulting size distribution in
an \emph{evolved} disk at several distances from the star.
It shows the case of a mixture of 70\% ice and 30\% astrosil, but the results
for the two other compositions are qualitatively similar.
We start a discussion of it with the parent ring (55--$90\AU$).
For grains larger than $s_\mathrm{c} \approx 10\mum$, the differential size distribution
has a slope $q \approx 3.7$. Since Fig.~\ref{fig: distribution} (left) plots
the optical depth per size decade, this corresponds to a slope of $q - 3 = -0.7$.
This is close to what is expected theoretically
for collision-dominated disks \citep{dohnanyi-1969}.
However, from $s_\mathrm{c} \approx 10\mum$ down to the smallest grains, the size distribution
flattens, because the inward drift is faster for smaller grains~-- or, more exactly,
for grains with higher $\beta$-ratios
\citep[see, e.g.,][for a more detailed discussion of this phenomenon]%
{strubbe-chiang-2006,vitense-et-al-2010}.
The slope calculated analytically for a transport-dominated disk
with $\beta \propto s^{-1}$ is $q \approx 2.5$.
The actual size distribution is wavy, and is rather close to $q = 3.0$,
which would correspond to a uniform distribution of the optical depth across the sizes.
One reason for that is a nonlinear dependence of $\beta$ on
the reciprocal of particle size (Fig.~\ref{fig: beta-size}).

Closer to the star, the slope for big grains progressively steepens
because of their preferential collisional elimination.
At the same time, the break $s_\mathrm{c}$ in the size distribution moves to smaller sizes.
Already at 20\AU, particles larger than $\sim 10\mum$ are almost absent.
At 10\AU, the cutoff shifts to $s \sim 4\mum$.
We argue that a progressive depletion
of larger grains with decreasing distance will be strengthened further 
by the alleged outer planet at $40\AU$.
That planet, which presumably sculpts the outer ring, would stop bigger grains more
effectively than smaller ones, by trapping some of them in mean-motion resonances (MMRs) and
scattering the others
\citep{liou-et-al-1996,moro-martin-malhotra-2003,moro-martin-malhotra-2005}.
As a result, we do not expect any grains larger than about $1$--$10\mum$
($\beta \la 0.1$--$0.01$) throughout most of the intermediate zone and in the entire inner region.
At the same time, the outer planet is not expected to be an obstacle for 
smaller particles that, as discussed below,  will be the most important 
for the mid-IR part of the SED.

From the same \emph{ACE} run for the outer and intermediate regions,
we found that $\tau$ is nearly constant
from 55\AU\ down to 10\AU\ (Fig.~\ref{fig: distribution} right).
This optical depth is dominated by submicron-sized
and micron-sized grains.
A constant optical depth inward from the sources is known to
be a characteristic feature of transport-dominated disks
\citep[e.g][]{briggs-1962,wyatt-2005b}.
However, disks with the optical depth of $\tau \ga 10^{-5}$ are usually
thought to be collision-dominated \citep{wyatt-2005b}, and one might not expect that even
the amplification of the Poynting-Robertson drag by stellar wind drag would
increase the critical optical depth (that separates the transport- and collision-dominated
regimes) by more than an order of magnitude.
This raises a natural question: how can it be that a disk
with the optical depth of $\sim 2 \times 10^{-4}$ is transport-dominated?
To find an answer, we note that $s_\mathrm{c}$ discussed above is a critical grain
size, for which the collisional lifetime is equal to the characteristic inward drift
timescale.
Therefore, the disk is collision-dominated at $s > s_\mathrm{c}$
and transport-dominated at $s < s_\mathrm{c}$.
In the \eqq{usual} disks, increasing $\tau$ would sooner or later force $s_\mathrm{c}$ to
reach the radiation pressure blowout limit $s_\mathrm{blow}$.
At that value of $\tau$, the entire disk becomes collision-dominated, since there is
very little material in the disk of size $s < s_\mathrm{blow}$.
But not in the $\varepsilon$~Eri disk!
Here, the blowout limit does not exist (Fig.~\ref{fig: beta-size}).
Therefore, for any reasonable size distribution ($q \ge 3$),
the optical depth is dominated by small particles with $s < s_\mathrm{c}$, and
these are in the transport-dominated regime. That is why our
simulation shows that the outer and intermediate regions of the $\varepsilon$~Eri disk is
dominated by grains roughly in the $0.05$--$1\mum$ size range and,
at those sizes, is nearly collisionless, despite $\tau$ of $\sim 10^{-4}$.
The same conclusion holds, of course, for the inner disk inside 10$\AU$.
For this reason, we believe our model of the inner disk in the following section,
which is obtained via collisionless numerical integration, is appropriate.

\section{Dust in the inner region} \label{sec: inner}

\subsection{Model}

To investigate the behavior of dust in the inner region ($r < 10\AU$)
of the $\varepsilon$~Eri debris disk and its interaction with the inner planet,
we performed numerical integrations of grain trajectories.
We used a Burlisch-Stoer algorithm \citep{press-et-al-1992}.
Collisions are not considered in our inner disk model, because, as shown above,
they play a minor role in the inner disk.

The dust grains were treated as massless particles, described only by their
$\beta$ ratio (or sizes) and their orbital elements:
semimajor axis $a$,
eccentricity $e$,
inclination $i$,
longitude of ascending node $\Omega$,
argument of pericenter $\omega$,
and mean anomaly $M$.

In our simulations we examined the following grain sizes $s_i$ ($i=1,...5$):
$0.07$, $0.11$, $0.43$, $0.84$, and $2.42\mum$.
These correspond to the $\beta$ ratios of $\beta_{i}=0.1$, 0.2, 0.2, 0.1, and 0.03,
respectively (Fig.~\ref{fig: beta-size}).
Each $s_i$ represents a size interval [$\breve{s}_{i}, \hat{s}_{i}$].
The limits of these intervals are set to the middle of $s_{i}$
and the adjacent sizes $s_{i-1}$ and $s_{i+1}$ (in logarithmic scale).
Since, due to the low luminosity of the central star,
the blowout grain size does not exist,
the lower cut-off for the particle sizes
was arbitrarily set to $\breve{s}_{1} = 0.05\mum$.
Smaller particles are not expected to contribute significantly to the
SED at mid-IR and longer wavelengths.
In addition, we expect that various erosive effects
(e.g., plasma sputtering) and dynamical effects (e.g., the Lorentz force),
which are not included in our model,
would swiftly eliminate the tiniest grains from the system.
Particles larger than $\hat{s}_{5} = 4\mum$ are not considered in these simulations,
because they are absent in this region (cf. Fig.~\ref{fig: distribution}, right).

Since grains with the same $\beta$ ratios experience the same force, it was enough
to run three simulations ($\beta_{i}=0.2$, 0.1, and 0.03) to cover all five grain
sizes.
We used 10\,000 particles for each value of $\beta$.
The grains were placed in orbits with
initial eccentricities from $0$ to $0.3$ and
an initial semimajor axis of 20\AU{}\footnote{We use $20\AU$
instead of $10\AU$ because, if we placed the particles at
$10\AU$ with eccentricities from $0$ to $0.3$, the initial distances would
be distributed between $7$ and $13\AU$, causing an unwanted \eqq{boundary effect}
of decreased optical depth in that distance range.},
assuming that they have passed the expected outer planet and
are out of the range of its perturbation.
The rather high initial eccentricities of up to $0.3$ were taken, because
these are expected to be increased by perturbations
of the outer planet when passing through its orbit.
The initial inclinations were uniformly distributed
between 0\grad and 10\grad, respectively.
The angular elements ($\Omega, \omega, M$) were all uniformly distributed
between 0\grad and 360\grad.

For the inner planet we used two different orbital configurations.
These configurations fit the observations from \citet{benedict-et-al-2006} and
\citet{butler-et-al-2006}, in the following called cases \eq{A} and \eq{B}, respectively.
In both cases we set the planet's semimajor axis to 3.4\AU{}
and assumed its orbit to be co-planar with the disk's mid-plane.
\citet{benedict-et-al-2006} estimate the planet's mass and eccentricity
to $\mass_{P_{A}} = 0.78\mass_\mathrm{Jup}$ and $e_{P_{A}} = 0.7$, while
\citet{butler-et-al-2006} give $\mass_{P_{B}} = 1.06\mass_\mathrm{Jup}$ and $e_{P_{B}} = 0.25$.
We also produced a simulation without an inner planet.
A summary of the initial parameters for the simulations is given in
Table~\ref{tab: simulation parameters exozodi}.

\begin{table}[th!]
   \caption{
   Initial parameters of the inner disk simulations
   } \label{tab: simulation parameters exozodi}
  \centering
  \begin{tabular}{lcc}
  \toprule
  \midrule
  star    & & \\
  \midrule
  $\mass_{\star}$   & \multicolumn{2}{c}{$0.83\mass_{\odot}$} \\
  $L_{\star}$       & \multicolumn{2}{c}{$0.32L_{\odot}$} \\
  $\dot{\mass}_{\star}$ & \multicolumn{2}{c}{$30\dot{\mass}_{\odot} = 6\times 10^{-13}\mass_{\odot}\yr^{-1}$} \\
  $v_{sw}$          & \multicolumn{2}{c}{$400\kmpersec$} \\
  \midrule
  \midrule
  planet    & A & B \\
  \midrule
    $\mass_{P}$ & $1.78 \times 10^{-3}\mass_{\star}$ & $2.44 \times 10^{-3}\mass_{\star}$ \\
    $a_{P}$ & 3.4\AU & 3.4\AU \\
    $e_{P}$ & 0.7 & 0.25 \\
    $i_{P}$ & 0.0\grad & 0.0\grad \\
    $M_{P}, \Omega_{P}, \omega_{P}$ & 0.0\grad & 0.0\grad \\
  \midrule
  \midrule
    \multicolumn{3}{l}{dust} \\
  \midrule$\beta$  & \multicolumn{2}{c}{0.2, 0.1, 0.03, 0.01} \\
    $a_0$      & \multicolumn{2}{c}{20\AU} \\
    $e_0$      & \multicolumn{2}{c}{[0.0, 0.3]} \\
    $i_0$      & \multicolumn{2}{c}{[0\grad, 10\grad]} \\
    $M_0, \Omega_0, \omega_0$      & \multicolumn{2}{c}{$[0\grad, 360\grad]$} \\
  \midrule
  \bottomrule
  \end{tabular}
\begin{flushleft}
  {\footnotesize A range of [$a, b$] indicates that the parameters are equally distributed between $a$ and $b$, and randomly chosen.}
\end{flushleft}
\end{table}

\subsection{Results}

\begin{figure}[thb!]
  \centering
  \includegraphics[width=0.48\textwidth]{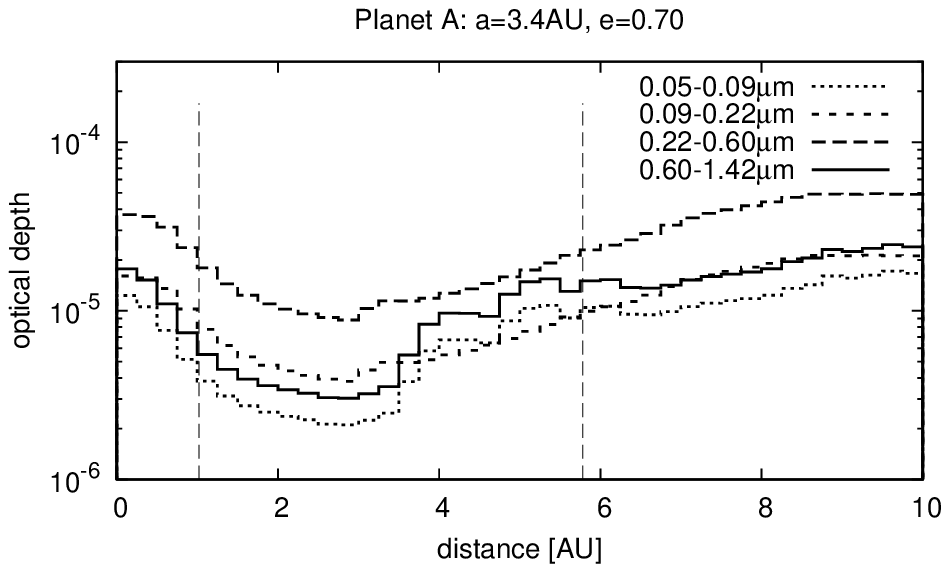} \\
  \includegraphics[width=0.48\textwidth]{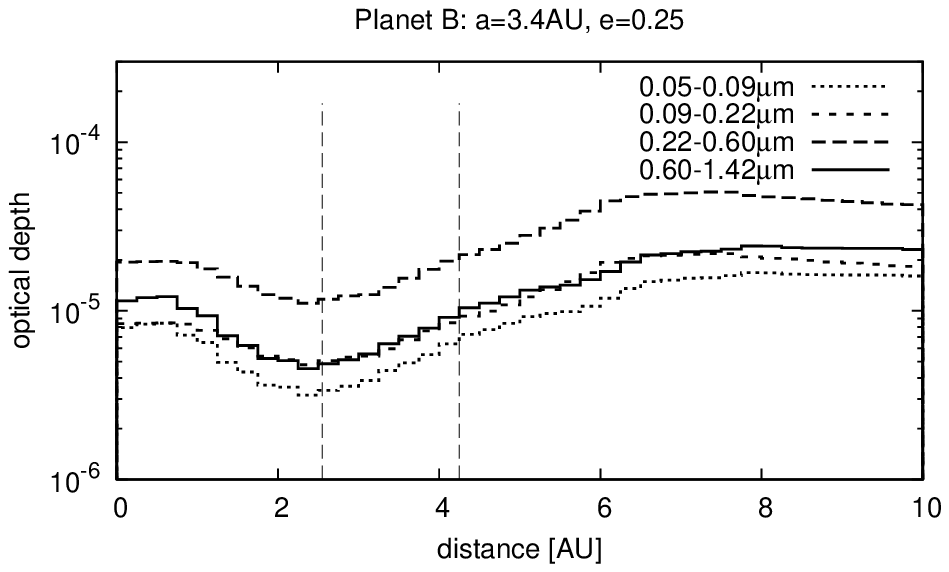}   \\
  \includegraphics[width=0.48\textwidth]{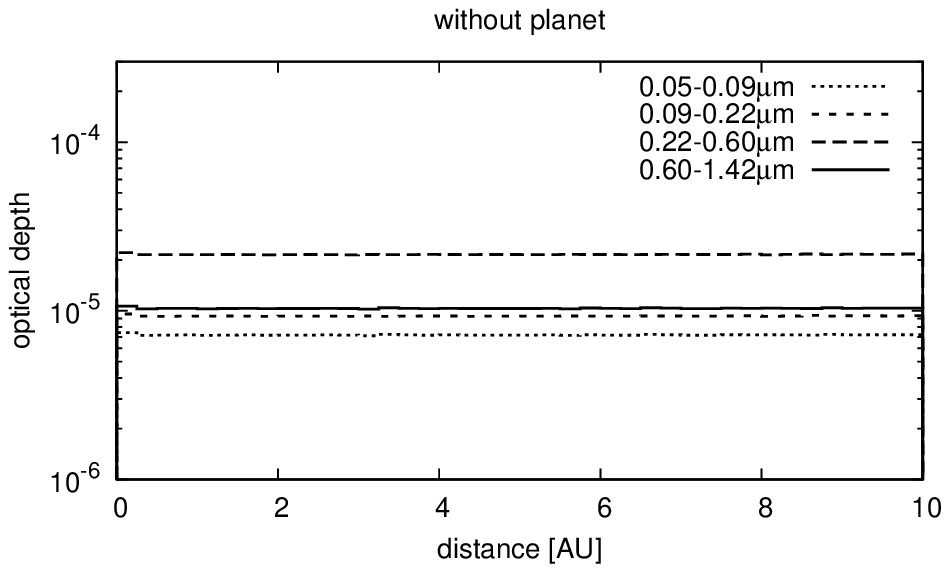}
  \caption{
    Radial profile of the normal geometrical optical depth
    produced by different grain size bins
    for the planet `A' ({\em top}),
    `B' ({\em middle}),
    and without a planet ({\em bottom}).
    A radial bin width is $\Delta r = 0.25\AU$.
    Vertical dashed lines show periastron and apastron of the planet orbit.
    A size distribution with a slope of $q=3.0$ and the dust mass
    of $\mass_\mathrm{dust} \approx 8.5 \times 10^{-8}\mass_{\oplus}$ (case `A'),
    $9.9 \times 10^{-8}\mass_{\oplus}$ (case `B'),
    and $5.6 \times 10^{-8}\mass_{\oplus}$ (without a planet)
    were used for the vertical scaling of the curves.
  }
  \label{fig: optical depth}
\end{figure}

From our simulations we created a radial profile
of the normal geometrical optical depth, assuming a steady-state
distribution (Fig.~\ref{fig: optical depth}).
It illustrates that the inner planet intercepts a fraction of dust on its way inward
by capturing the grains in MMRs and/or scattering those grains that pass the planetary orbit.
This results in an inner gap around and inside the planet orbit.
The gap is deeper for grains with smaller $\beta$ ratios, because they
drift inward more slowly, which increases the probability of a resonance trapping or
a close encounter with the planet.
Another conclusion from Fig.~\ref{fig: optical depth} is that the radial profiles
in the cases `A' and `B' do not differ much, although the eccentricities
of planetary orbits are quite different.

\section{Spectral energy distribution} \label{sec: SED}

\subsection{SED from the outer and intermediate regions} \label{ssec: sed outer}
Having calculated the distribution of dust particles with different $\beta$ ratios
in the $\varepsilon$~Eridani system, we need to calculate the
thermal emission of the dust for comparison with the observed SED.
We start with the SED of the outer and intermediate region together.
We have calculated it, using the output of the \emph{ACE} runs described in
Sect.~\ref{sec: outer_interm}.
Specifically, coupled radial and size distributions
of dust between 10 and 90\AU{} shown in Fig.~\ref{fig: distribution} were used.

\begin{figure}[t]
  \centering
  \includegraphics[width=0.48\textwidth]{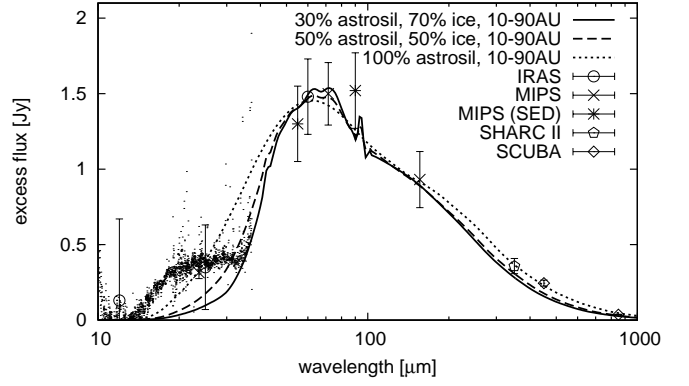} \\
  \caption{
   The SED from dust outside $10\AU$ in the $\varepsilon$~Eri dust disk.
   Symbols with error bars are data points and scattered dots are the IRS spectrum.
   Lines are model SEDs for different dust compositions:
   100\% astrosil (dotted),
   50\% ice and 50\% astrosil (dashed), and
   70\% ice and 30\% astrosil (solid).
}
    \label{fig: sed_outer}
\end{figure}

The results are shown in  Fig.~\ref{fig: sed_outer}.
Overplotted are the data points,
all taken from Fig.~7 of \citet{backman-et-al-2009};
see also their Table~1.
Clearly, the model with 100\% astrosilicate contents is too \eqq{warm}.
Notably, the SED starts to rise towards the main maximum too early
(at $\approx 25\mum$ instead of $\approx 35\mum$),
which is inconsistent with the \eqq{plateau} of the IRS spectrum.
Including 50\% ice improves the model,
but the SED still rises too early.
The best match of the data points is achieved with a 30\% astrosil and 70\% ice mixture,
which will therefore be used in the rest of the paper.
Still, even for that mixture, the main part of the SED (i.e. the combined
contribution of the outer ring and the intermediate region) predicted by the collisional
model is somewhat \eqq{warmer} than it should be.
We discuss this slight discrepancy in Sect.~\ref{ssec: sed}.

\subsection{SED from the inner region} \label{ssec: sed of the inner region}

We now consider an SED coming from dust in the inner disk.
The optical constants, density of the dust, and the stellar spectrum described
in Sect.~\ref{ssec: grain sizes} have been used to calculate the \eqq{partial} SEDs
from each grain size bin.
To sum up the contributions from different bins, we assumed a power law
$\mass(s)\diff{s} \propto s^{3-q}\diff{s}$.
There are two parameters that allow us to adjust the resulting SED of the inner region
to the IRS spectrum.
One is the slope $q$ that
provides a \eqq{weighting} of relative contributions of
different-sized grains into the resulting SED.
In accord with our results in Sect.~\ref{ssec: results},
we take $q=3$ and only consider grains with $s \la 2\mum$
(four lowest size bins).
Another parameter should characterize the absolute amount of
dust in the inner region. (This could be, for instance,
the total dust mass in the inner region.)
This parameter would determine the overall height of the SED.
We varied the dust mass until the resulting model SED fits the IRS
spectrum best.
The best results were achieved with
the total dust mass in the inner region
of $\mass_\mathrm{dust} \approx 8.5 \times 10^{-8}\mass_{\oplus}$ in the case `A',
$9.9 \times 10^{-8}\mass_{\oplus}$ in the case `B',
and $5.6 \times 10^{-8}\mass_{\oplus}$ without a planet.
The absolute height of optical depth profiles from various size bins
shown in  Fig.~\ref{fig: optical depth}
corresponds to the same dust masses and the same slope $q=3.0$.

\begin{figure}[thb!]
  \centering
  \includegraphics[width=0.48\textwidth]{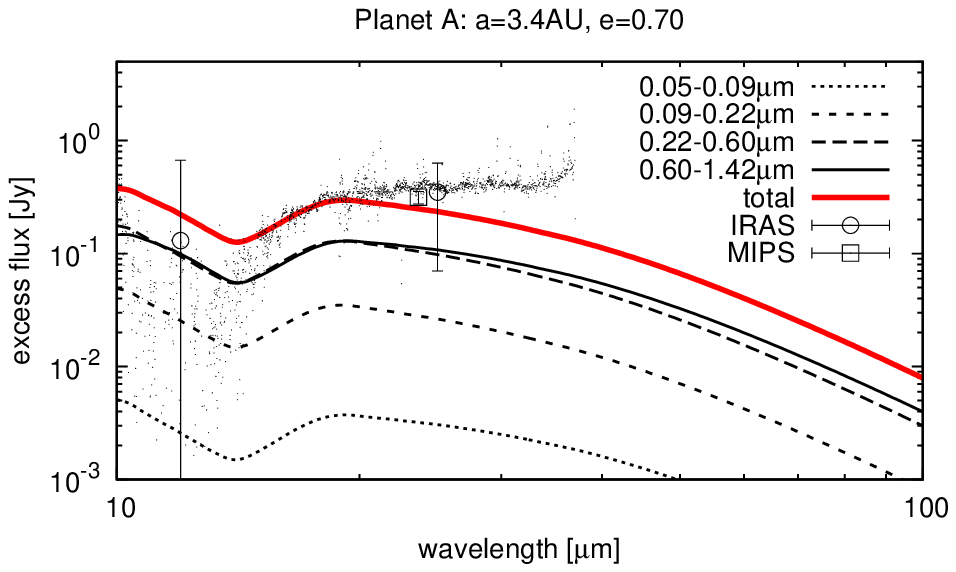} \\
  \includegraphics[width=0.48\textwidth]{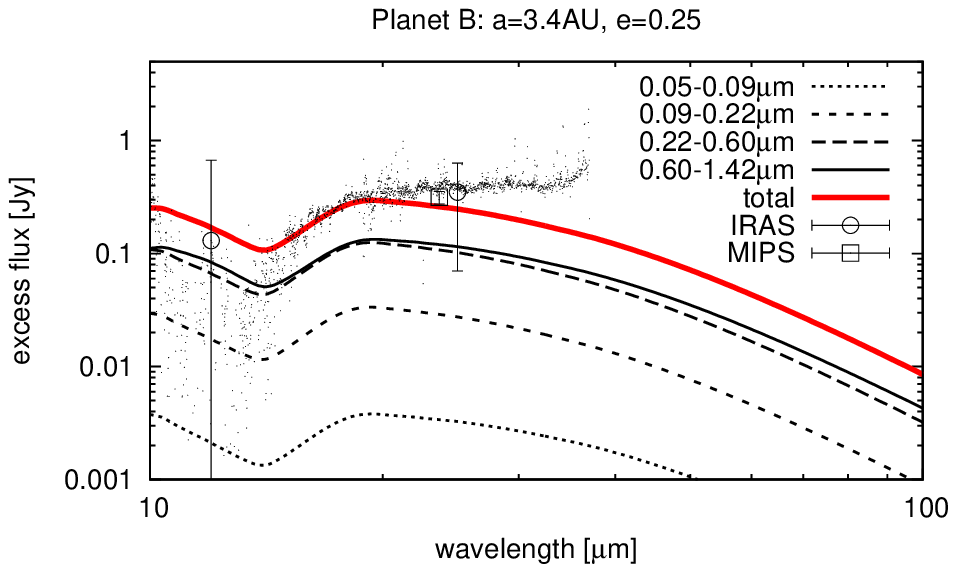}   \\
  \includegraphics[width=0.48\textwidth]{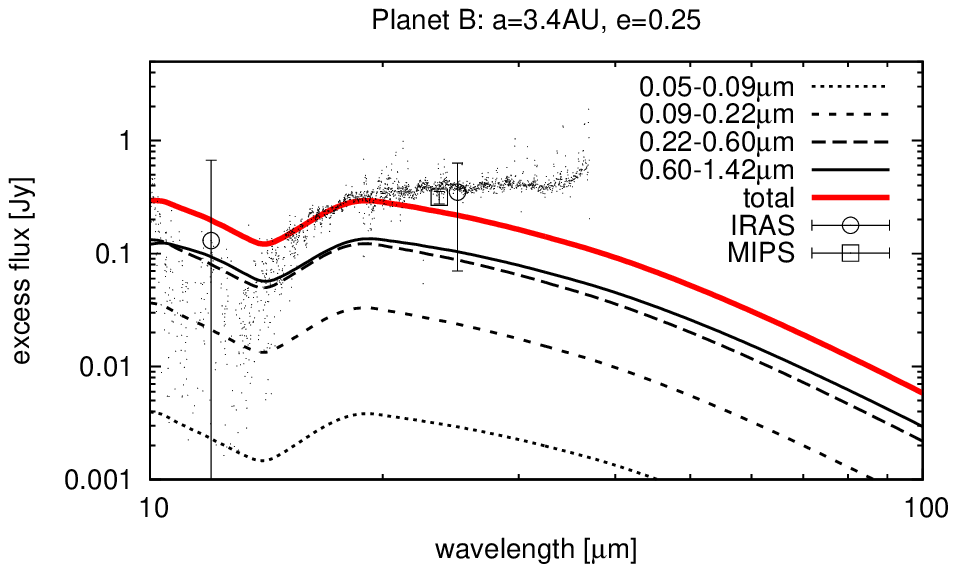}
  \caption{
    The observed Spitzer/IRS spectrum \citep{backman-et-al-2009} (small dots)
    and the modeled SED of the inner region.
    {\em Top:} planet `A',
    {\em middle:} planet `B',
    {\em bottom:} without a planet.
    Thin lines: contributions of the different size bins to the SED,
    thick line: the total emission from all sizes.
    The size distribution slope $q$ and the dust mass
    $\mass_\mathrm{dust}$ are the same as in Fig.~\ref{fig: optical depth}.
  }
  \label{fig: sed_inner}
\end{figure}

Figure~\ref{fig: sed_inner} shows the contribution of the different grain sizes 
to the SED and their total. we now use a logarithmic scale.
Although the two planet orbits are quite different, the
influence of the planet on the SED is rather minor,
because the radial distributions of dust are similar in both cases.

A consistency check that we made was to compare the model predictions
with the results of interferometric measurements with CHARA array
in the $K$-band (2.2\mum) \citep{difolco-et-al-2007}.
They set the upper limit of the fractional excess emission of
the inner debris disk to $6 \times 10^{-3}$ ($3\sigma$ upper limit).
With the photospheric flux of 120\Jy{} at $\lambda = 2.2\mum$,
this translates to an excess of $\la 720\mJy$. This
value includes both thermal emission and scattered light.
The integrated surface brightness of the 2.2\mum{} radial thermal emission
profile, convolved with the CHARA transmission profile, generates a total
excess of just 25.7\mJy, 14.5\mJy, and 15.4\mJy{}
for the cases \eq{A}, \eq{B}, and without an inner planet, respectively.
Even if we took scattered light into account, which we estimate 
to contribute $\approx 3.4$ times more than the thermal emission at that wavelength,
our model would be consistent with non-detection of dust with CHARA. 

\subsection{SED from the entire disk} \label{ssec: sed}

\begin{figure}[t]
  \centering
  \includegraphics[width=0.48\textwidth]{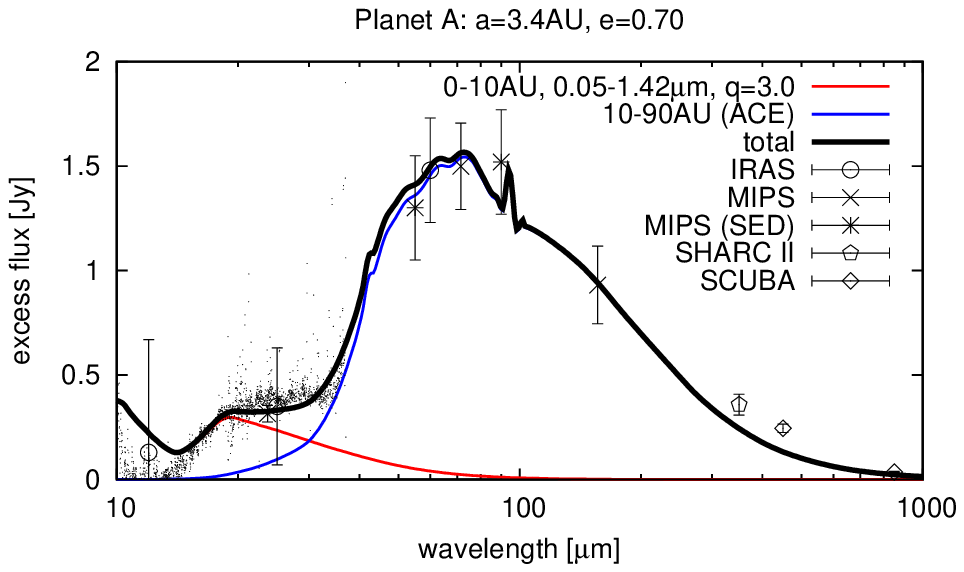} \\
  \includegraphics[width=0.48\textwidth]{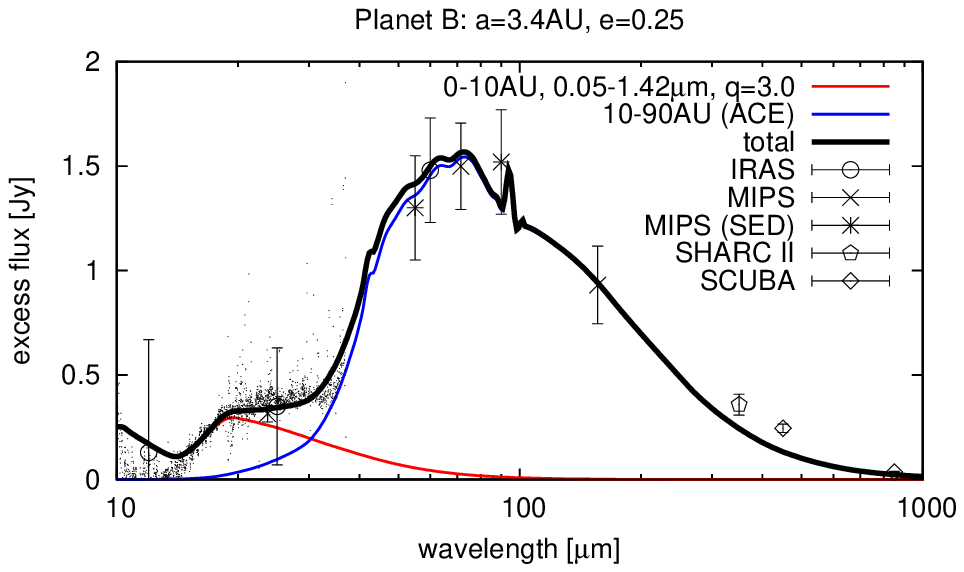} \\
  \includegraphics[width=0.48\textwidth]{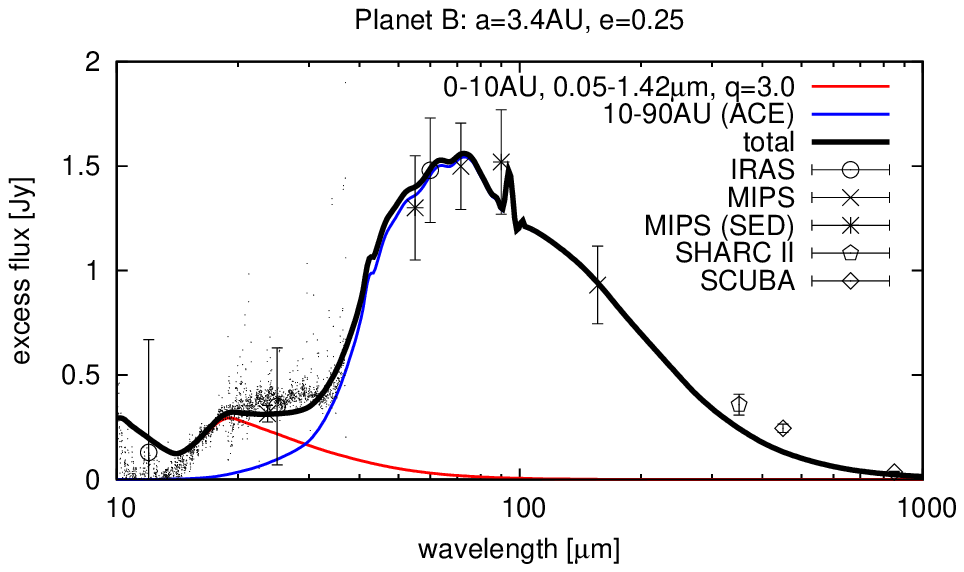}
  \caption{
   The entire SED of the $\varepsilon$~Eri dust disk.
   Symbols with error bars are data points,
   small dots is the IRS spectrum.
   Thick line in each panel is the excess emission predicted by the models,
   and thin lines show contributions from the inner region ($<10\AU$)
   and the intermediate$+$outer disk.
   Three panels are for the planet configurations \eq{A} ({\em top}),
   \eq{B} ({\em middle}), and without a planet ({\em bottom}).
}
    \label{fig: sed}
\end{figure}

We now assemble the SED produced by the entire disk.
To this end,
we summed up the SEDs of the inner region and of the region outside $10\AU$
presented in Figs.~\ref{fig: sed_outer}~and~\ref{fig: sed_inner}, respectively.
Figure~\ref{fig: sed} shows the complete SED.
It is in a reasonable, although not a perfect,
agreement with the observations.
In particular, the maximum of the modeled SED, while reproducing
the data points within their error bars, appears to lie at a slightly
shorter wavelength than the one suggested by the data points.
A likely reason for this discrepancy is that our collisional
simulation does not take into account elimination of particles in the size range
from $\sim 1$ to $\sim 100\mum$ by the alleged outer planet,
as explained above. 
Excluding these particles from the intermediate region 10--55\AU{}
would reduce emission in the 35--70\mum{} wavelength range, shifting
the maximum of the SED to a longer wavelength.
In addition, the main part of the SED can be made \eqq{colder}
by varying diverse parameters of the collisional
simulation, many of which are not at all or are poorly constrained.
These include the eccentricity distribution of planetesimals, the opening angle
of the planetesimal disk, as well as the mechanical strength of solids.
Such a search for the best fit would, however, be very demanding computationally.
We deem the fit presented in Fig.~\ref{fig: sed}
sufficiently good to demonstrate that our scenario, in which inner warm dust is produced
in the outer ring, is feasible.

\subsection{Connecting outer and inner regions} \label{ssec: connection}

To make sure that the amount of dust required to reproduce the IRS
spectrum is consistent with the amount of dust that could be supplied to the inner disk
from the outer parent belt,
we now make an important consistency check.
From calculations of the dust production in the outer ring and its transport
through the intermediate region, we know the optical depth at $10\AU$.
Figure~\ref{fig: distribution} (right) suggests that the optical depth per size decade,
created by grains with sizes of $0.07$, $0.11$, $0.43$, and $0.84\mum$, 
is about (1--2)$\times 10^{-4}$.
These values have to be compared with optical depths,
sufficient to reproduce the IRS spectrum.
From Fig.~\ref{fig: optical depth} outside the planetary region we read out
the values (2--5)$\times 10^{-5}$ for the four lowest size bins
(which are centered on the same four grain sizes).
Taking into account the actual width of the four lowest size bins used in the
modeling of the inner system, which is $\approx 0.4$ dex,
we would need an optical depth per size decade of (5--10)$\times 10^{-5}$.
Thus the optical depth that is supplied by grains transported from outside
matches the optical depth required to account for the IRS spectrum within a factor of two.

\section{Surface brightness profiles} \label{sec: profiles}

The model developed in this paper was designed to explain the Spitzer/IRS
spectrum of the system. Besides this, we have made sure that it reproduces the entire
SED probed with many instruments at the far-IR and sub-mm wavelengths.
We now provide a comparison with other measurements that we have not considered before.

The most important information comes from spatially resolved images.
Spitzer/MIPS observations yielded brightness profiles
in all three wavelength bands centered on 24, 70, and $160\mum$
\citep{backman-et-al-2009}. We calculated the brightness profiles
of thermal emission with our model.
In doing so, we included both the outer+intermediate disk and the inner one.
The resulting brightness profiles were convolved with the instrumental PSF
\citep[see Sect.~2 in][for the algorithm used]{mueller-et-al-2010}
and compared them with observed profiles.

\begin{figure}[t!]
  \centering
  \includegraphics[width=0.48\textwidth]{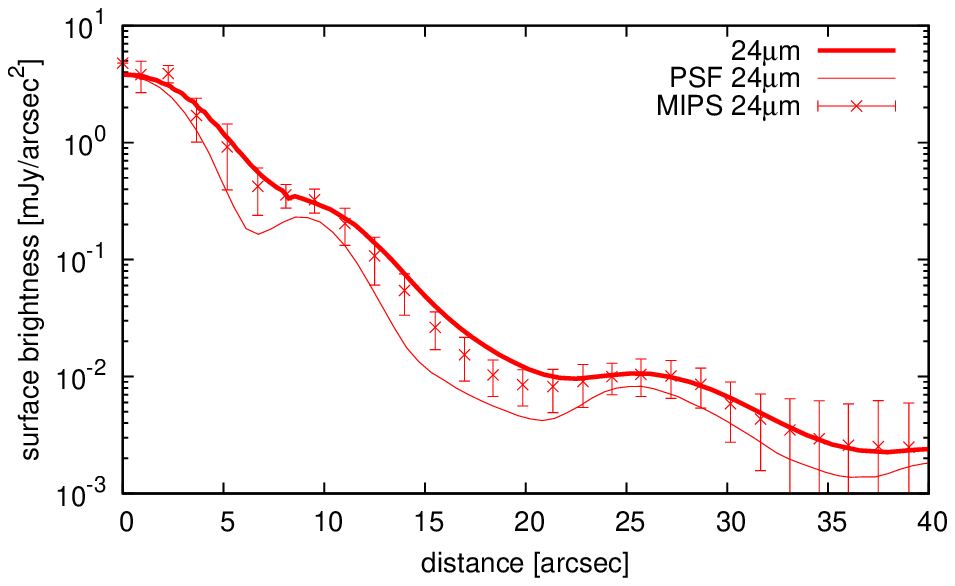} \\
  \includegraphics[width=0.48\textwidth]{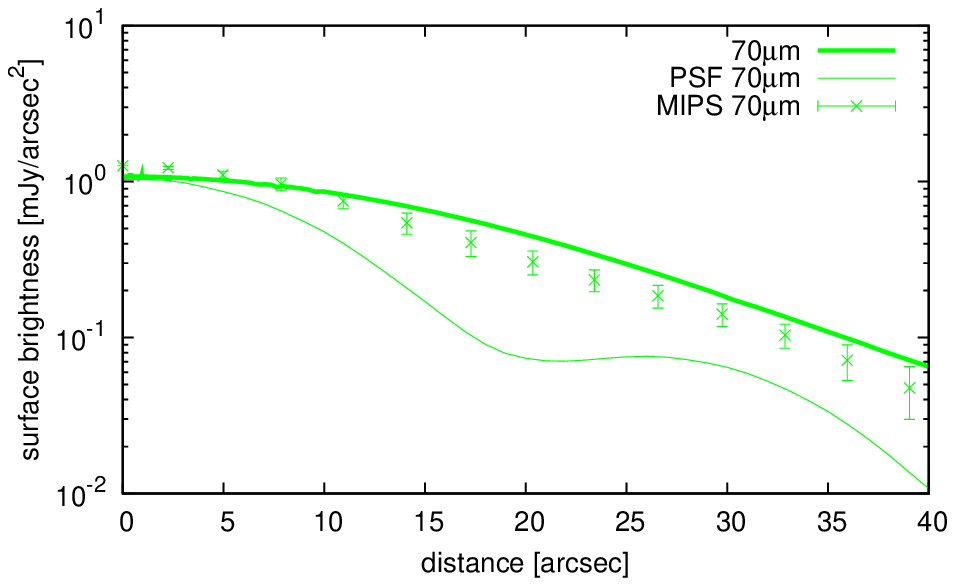} \\
  \includegraphics[width=0.48\textwidth]{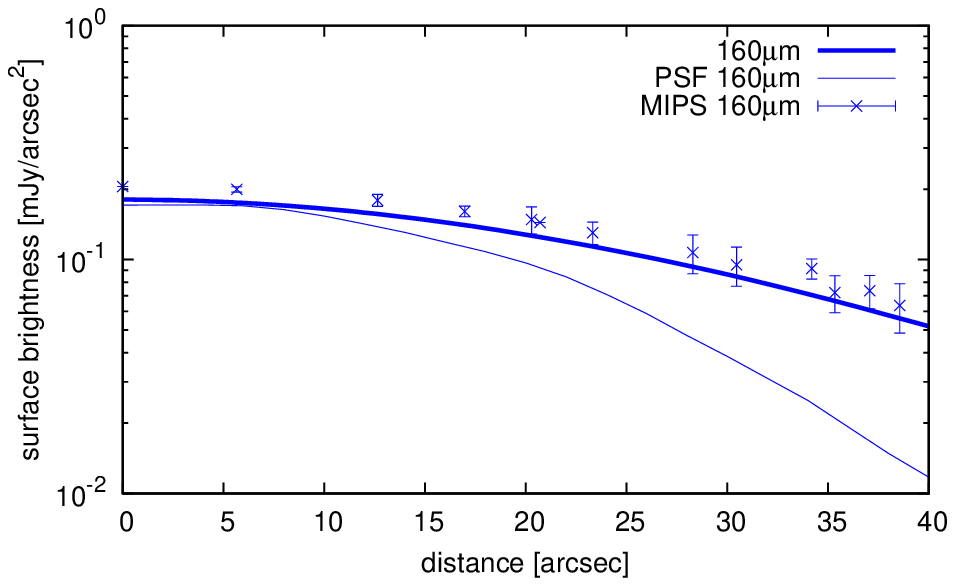} \\
  \caption{
   Azimuthally averaged surface brightness brightness profiles
   of the $\varepsilon$~Eri dust disk (case `A')
   at three Spitzer/MIPS wavelengths:
   $24\mum$ ({\em top}),
   $70\mum$ ({\em middle}),
   and $160\mum$ ({\em bottom}).
   Points with error bars: star-subtracted MIPS data
   \citep[Fig. 6 of][]{backman-et-al-2009}.
   Solid lines: modeled profiles after convolution with the PSFs.
   Thin solid lines: the instrumental PSF.
}
    \label{fig: profiles}
\end{figure}

The results are presented in Fig.~\ref{fig: profiles}.
We only show case `A', since the profiles in case `B' and without
a planet are very similar.
At all three wavelengths, the modeled brightness monotonically increases
toward the star, as does the observed brightness. Both the slopes
and the absolute brightness level are in good agreement with observations.
The only exception is the modeled $70\mum$ profile (Fig.~\ref{fig: profiles} middle).
It is flatter than the observed one, predicting the brightness inside
$30\AU$ correctly but overestimating the emission
in most of the intermediate and the parent ring regions by a factor of two.
The reasons for this deviation are probably the same as those discussed
in Sect.~\ref{ssec: sed}.
First, if a presumed outer planet at $\approx 40\AU$ efficiently
eliminates particles in a size range from several to several tens of micrometers
(which is not taken into account in our {\emph ACE} simulations),
this will decrease the $70\mum$ brightness in the intermedate region.
Second, it should be possible to slightly decrease
the $70\mum$ emission in the parent ring region by
varying poorly known parameters in the collisional simulation.

We also checked the radial brightness profile at $850\mum$ and compared
it with JCMT/SCUBA observations \citep{greaves-et-al-1998,greaves-et-al-2005}.
After convolution with a Gaussian PSF of $\sigma = 5''$, the resulting profile
is consistent with \citet[][their Fig.~2]{greaves-et-al-2005},
showing a broad ring around $\approx 60\AU$ and a resolved central cavity.

\section{Conclusions} \label{sec: conclusions}

In this paper we show that in the nearby system $\varepsilon$~Eridani the
Spitzer/IRS excess emission at $\lambda \sim 15$--$30\mum$ can be caused by dust that
is produced in the known outer dust ring and that streams inward due to
interaction with strong stellar winds.

By running a collisional code, we simulated the dust production in the outer ring
between 55\AU{} and 90\AU{} with a dust mass of $10^{-3}\mass_{\oplus}$
and the subsequent transport of the dust inward to 10\AU{}.
We then employed single-particle numerical integrations to simulate the dust transport
further inward through the orbit of the known inner planet.
The dust in the inner region was found to consist of grains
smaller than $\approx 2\mum$, and the dust mass inside $10\AU$ was estimated to be
(6--10)$\times 10^{-8}\mass_{\oplus}$.
Combining the results of the collisional simulations outside $10\AU$
and numerical integrations inside that distance, we calculated the
overall SED and radial brightness profiles.
This SED is in a reasonable agreement with the available observational
data, and it correctly reproduces the shape and the height of the Spitzer/IRS spectrum.
Likewise, the brightness profiles are consistent with the Spitzer/MIPS data.

The best results are obtained with an ice-silicate composition
\citep{laor-draine-1993, li-greenberg-1998-03} of dust outside the
ice sublimation distance of $\approx 10\AU$, and an inner disk
of non-volatile silicate grains inside that distance.

With the aid of the modeled spectra and brightness profiles, it is not possible to distinguish
between the different orbital solutions for the inner, radial velocity planet
proposed by \citet{benedict-et-al-2006} or \citet{butler-et-al-2006}. Although
the planetary orbits they inferred are quite different, both setups yield
quite similar radial profiles of dust, the SEDs, and the brightness profiles.

Various kinds of new data on the $\varepsilon$~Eri system are
expected soon. Data from JCMT/SCUBA2 (J. Greaves, pers. comm.),
HERSCHEL/PACS, and /SPIRE should shed light on the cold dust.
This includes the structure of the outer ``Kuiper belt''
and the intermediate region of the disk between 10 and 55\AU{}.
Further more, there is hope to better probe the inner warm dust directly,
with instruments such as the Mid InfraRed Instrument (MIRI)
aboard the upcoming James Web Space Telescope (JWST).
Likewise, there is an ongoing effort to
find outer planets in the system by direct imaging
\citep{itoh-et-al-2006,marengo-et-al-2006,janson-et-al-2007,janson-et-al-2008,marengo-et-al-2009}.

\acknowledgements
  We thank Johan Olofsson for assistance with reducing the Spitzer/IRS spectrum
  and Massimo Marengo and Karl Stapelfeldt for providing us with
  the Spitzer/MIPS brightness profiles.
  Useful discussions with Hiroshi Kobayashi are acknowledged.
  We appreciate critical readings of the manuscript by Philippe Th\'ebault
  and by the anonymous reviewer.
  Part of this work was supported by the
  \emph{Deut\-sche For\-schungs\-ge\-mein\-schaft (DFG)}, projects
  Kr~2164/8--1 and Kr~2164/9--1, by the \emph{Deutscher Akademischer
  Austauschdienst (DAAD)}, project D/0707543, and by the International
  Space Science Institute in Bern, Switzerland (\eqq{Exozodiacal Dust Disks
  and Darwin} working group\footnote{\sf http://www.issibern.ch/teams/exodust/}).
  SM was funded by the graduate student fellowship of the Thuringia State.

%

%

\end{document}